\documentclass[journal]{IEEEtran}

\ifCLASSINFOpdf
\else
\fi

\usepackage{url}
\usepackage{amssymb}
\usepackage{algorithm}
\usepackage{cite}
\usepackage[noend]{algorithmic}
\usepackage{amsmath}
\usepackage{amssymb}
\usepackage{array}
\usepackage{comment}
\usepackage{graphicx}
\usepackage{pifont}
\usepackage{epsfig,tabularx,amssymb,amsmath,subfigure,multirow,booktabs}
\usepackage{balance}
\usepackage{mathptmx}
\usepackage{caption2}

\newtheorem{theorem}{Theorem}

\newtheorem{definition}{Definition}
\begin{document}
%
% paper title
% Titles are generally capitalized except for words such as a, an, and, as,
% at, but, by, for, in, nor, of, on, or, the, to and up, which are usually
% not capitalized unless they are the first or last word of the title.
% Linebreaks \\ can be used within to get better formatting as desired.
% Do not put math or special symbols in the title.
\title{Embedding the Minimum Cost SFC with End-to-end Delay Constraint}

\author{
\IEEEauthorblockN{Bangbang Ren,  Ying Han}

\IEEEauthorblockA{National University of Defense Technology}
}

%% The paper headers
\markboth{Journal of \LaTeX\ Class Files,~Vol.~14, No.~8, August~2015}%
{Shell \MakeLowercase{\textit{et al.}}: Bare Demo of IEEEtran.cls for IEEE Journals}
% The only time the second header will appear is for the odd numbered pages
% after the title page when using the twoside option.
%
% *** Note that you probably will NOT want to include the author's ***
% *** name in the headers of peer review papers.                   ***
% You can use \ifCLASSOPTIONpeerreview for conditional compilation here if
% you desire.

\maketitle

\begin{abstract}
Many network applications, especially the multimedia applications, often deliver flows with high QoS, like end-to-end delay constraint. Flows of these applications usually need to traverse a series of different network functions orderly before reaching to the host in the customer end, which is called the service function chain (SFC). The emergence of network function virtualization (NFV) increases the deployment flexibility of such network functions. In this paper, we present heuristics to embed the SFC for a given flow considering: i) bounded end-to-end delay along the path, and ii) minimum cost of the SFC embedding, where cost and delay can be independent metrics and be attached to both links and nodes. This problem of embedding SFC is NP-hard, which can be reduced to the Knapsack problem. We then design a greedy algorithm which is applied to a multilevel network. The simulation results demonstrate that the multilevel greedy algorithm can efficiently solve the NP-hard problem.
\end{abstract}

% Note that keywords are not normally used for peerreview papers.

\IEEEpeerreviewmaketitle

\section{INTRODUCTION}
\label{introduction}
Many network applications deliver flows with high QoS, like the end-to-end delay constraint. With the popularity of mobile computing, multimedia applications are in a significant state of change that shifts user subscription patterns\cite{gerdfeldter2015setting,ren2018minimum}. For example, Anvato \cite{anvato} can make online video editing for content providers, ad insertion for advertisers, caching, and transcoding for heterogeneous user devices. These particular requirements for video flows are usually satisfied by traversing different network function middleboxes, which is called service function chain (SFC). Network Function Virtualization (NFV)\cite{NFVsurvey} is an emerging new technology to deploy network functions in software, which can bring benefits. The popular stream media firm, Netflix \cite{Netflix}, has adopted AWS to support their service chains with NFV.

However, the software-implementation of network functions through virtualization technologies on general servers may bring performance degradation, compared to the corresponding physical version on dedicated hardware~\cite{HanGJL15}. The reason is that the operation of Virtual Network Functions (VNFs) may be affected by surges in computing load, hardware malfunctions. This reason suggests that deploying VNFs in different VMs may suffer different delays and costs. The placement of VNFs will further influence the flow transmission cost. For any given flow, we want to embed the expected SFC with the minimum cost and satisfying the end-to-end delay constraint.
\begin{figure}[t]%
\centering
\includegraphics[scale=0.38]{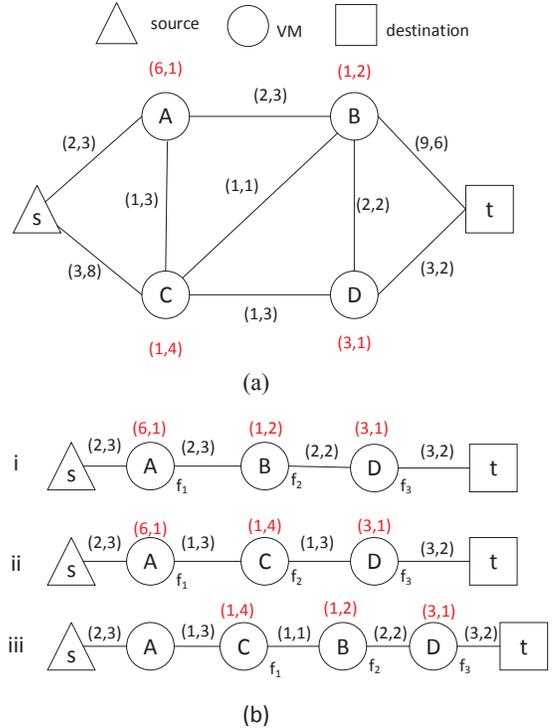}
\caption{(a) The original network with all links and nodes, each of which is attached a two-tuple weight, (cost,delay). (b)Three solutions of embedding the SFC $(f_1\rightarrow f_2\rightarrow f_3)$, in which each $f_i$ denotes a network function.}%
\vspace{-0.6in}
\label{exampleshow}%
\end{figure}

Fig. \ref{exampleshow}(a) shows a network, in which each edge and each node has a two-tuple weight, that is, (cost, delay). The two-tuple weights in edges are related to the flow transmission, while the weights in nodes concern the flow process. In this setting, when a VM does not deploy any VNF, the processing cost and delay are both $0$. There are four available VMs (A,B,C,D) to deploy the SFC $(f_1\rightarrow f_2\rightarrow f_3)$. In this paper, it is assumed that a VM can run at most one VNF. Indeed, the situation that VM can deploy multiple VNFs can be addressed by replicating the VM multiple times in the original network. Fig. \ref{exampleshow}(b) shows three feasible embedding paths. In solution i), the cost is 19 and the end-to-end delay is 14. In solution ii), the cost is 17 and the end-to-end delay is 17. While in solution iii), there are four VMs, but the VM A just act as an intermediate node which both the setup cost and processing delay are 0. Thus, the cost and end-to-end delay are 14 and 18, respectively. This observation demonstrates that there are different embedding solutions of the same SFC, which exhibit diverse total costs and the end-to-end delays. This paper aims to find an end-to-end delay guaranteed minimum cost SFC embedding (D-MCS) considering the node processing impacts and edge transmitting impacts.

\section{SYSTEM MODEL}
\label{systemmodel}

We consider a network $G=(V,E)$, where each link $e\in E$ is associated with a nonnegative cost $c_e$ denoting the connection cost and a nonnegative delay $d_e$ denoting the connection delay. Similarly, each virtual machine $v\in V$ has two nonnegative weight $c_v$ and $d_v$ which denote the setup cost and the data processing delay to run a VNF. Given a flow request from $s$ to $t$, whose the upper end-to-end delay bound is $\Delta$ and the required SFC requirement (a chain of VNFs) is $\ell\mathrm{=}(f_1,f_2,...,f_l)$. We devote to embedding the SFC into the network with the minimum cost while satisfy the end-to-end delay constraint. Here, we use a binary variable $x_{ij}$ to denote whether the link $(i,j)\in E$ is selected and $y_i$ to denote whether the node $i\in V$ is selected to deploy a VNF. We formulate this problem with an integer linear programming as follows:
%\vspace{-0.04in}
\begin{subequations}\label{opt}
\begin{align}
 &\underset{x_{ij},y_i}{\text{Min}}&& \sum_{(i,j)\in E}c_{ij}x_{ij} \mathrm{+} \sum_{i \in V}c_iy_i&  \label{eq:rel0}\\
 & &&\sum_{j\in~V}x_{ij}\mathrm{-}\sum_{k\in~V}x_{ki}=1, &i\mathrm{=}s \label{eq:rel1}\\
 & &&\sum_{j\in~V}x_{ij}\mathrm{-}\sum_{k\in~V}x_{ki}=-1, &i\mathrm{=}t \label{eq:rel2}\\
 & && \sum_{j\in~V}x_{ij}\mathrm{-}\sum_{k\in~V}x_{ki}=0, &i\in~V\backslash\{s,t\} \label{eq:rel3}\\
 &&&y_i\mathrm{=}0, &i\mathrm{=}s,t  \label{eq:rel4} \\
  &&&\sum_{(i,j)\in E}d_{ij}x_{ij}\mathrm{+}\sum_{i\in V}d_iy_i\mathrm{<}\Delta& \label{eq:rel5}\\
  &&&\sum_{i\in~V}y_i\mathrm{=}l& \label{eq:rel6}\\
  &&& y_i\mathrm{\leq}\sum_{j\in~V}x_{ij}, &\forall i\in V \label{eq:rel7}\\
  &&& x_{ij}\in~\{0,1\}, &\forall (i,j)\in E\label{eq:rel8}\\
  &&& y_i\in~\{0,1\} &\forall i\in~V \label{eq:rel9}
\end{align}
\end{subequations}

Eqn. (\ref{eq:rel0}) denotes our objective is to minimize the total cost of embedding the SFC, in which $c_{ij}$ and $c_i$ denote the cost of the links and nodes, respectively. Eqns. (\ref{eq:rel1}-\ref{eq:rel3}) model the flow constraint and ensure that the flow spreads along a path. Eqn. (\ref{eq:rel4}) put restricts on the source and the destination node that they cannot deploy VNF. Eqn. (\ref{eq:rel5}) ensures that the end-to-end delay will never violate the upper delay bound. Eqn. (\ref{eq:rel6}) requires that the number of nodes selected to deploy VNFs must be equal to the length of the SFC.

\begin{theorem}
\label{NPproof}
The above D-MCS problem is NP-hard.
\end{theorem}

\begin{figure}[h]%
\centering
\includegraphics[scale=0.38]{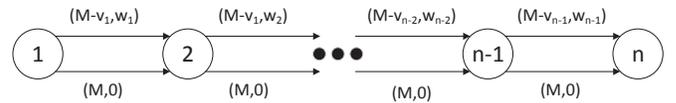}
\caption{An instance of D-MCS}%
\vspace{-0.02in}
\label{graphinstance}%
\end{figure}

\begin{IEEEproof}
We prove the NP-hardness of D-MCS by showing a reduction from the KNAPSACK problem \cite{Karp1975Reducibility}. An instance of KNAPSACK is: there are a knapsack and $(n-1)$ items. Each item has two attributes, a weight $w_i$ and a value $v_i$. The maximum weight of the knapsack can support is $\lambda$. Then we decide which items can be bagged into the knapsack so that the total weight is less than or equal to $\lambda$ and the total value is as large as possible. The integer programming of KNAPSACK is shown as follows.

\begin{subequations}\label{optknapsack}
\begin{align}
 &\underset{x_{j}}{\text{Max}}&& \sum^{n-1}_{j=1}v_jx_j&  \label{myrel0}\\
 & &&\sum^{n-1}_{j=1}w_jx_j\leq \lambda & \label{myrel1}\\
 & &&x_j \in \{0,1\} \quad j=1,...,n-1& \label{myrel2}
\end{align}
\end{subequations}

Now we construct an instance of D-MCS. In this instance, there are $n$ nodes and two parallel edges between each pair of neighbor nodes as shown in Fig. \ref{graphinstance}. The edges lies on the top have a two-tuple attribute $(M\mathrm{-}v_i,w_i)$, in which $M\mathrm{=}max(v_1,v_2,...v_{n-1})$, denoting the cost and delay of each edge. All nodes have a two-tuple attributes $(c_i,d_i)$. In the instance of D-MCS, we set $l\mathrm{=}1$, and the delay bound $\Delta\mathrm{=}\lambda + max(d_1,d_2,...,d_n)$. If we could get the optimal solution of the KNAPSACK problem, then we can find the path from $1$ to $n$ in Fig. \ref{graphinstance}. In detail, if an item is selected in the KNAPSACK problem, the corresponding top edge is included. Along the path, the sum delay is $\sum_{j=1}^{n-1} w_jx_j\leq \lambda$, and the sum cost is $\sum_{j=1}^{n-1} (M-w_j)x_j$. Since the KNAPSACK problem aims to maximize the total value $\sum_{j=1}^{n-1}w_jx_j$, then $\sum_{j=1}^{n-1} (M-w_j)x_j$ is the minimum value, that is, we could find the minimum cost path. We can compare all nodes to select the optimal node to deploy the VNF. Since the upper delay bound is $\Delta\mathrm{=}\lambda + max(d_1,d_2,...,d_n)$, any node selection will never violate the delay upper bound. That is if we could find a solution to this D-MCS problem instance, then the solution without deploying VNFs will be the optimal solution of the KNAPSACK problem. However, under the assumption that $P\neq NP$, this is impossible. Thus, D-MCS is NP-hard, Theorem \ref{NPproof} is proved.

\end{IEEEproof}

\section{ALGORITHM DESIGN}
In the following, we assume that the algorithm has prior knowledge about the whole network. The D-MCS requires that the path from $s$ to $t$ traverses at least $l$ nodes. Accordingly, we first construct a multilevel network with $l\mathrm{+}2$ levels, where the $2\mathrm{-}(l\mathrm{+}1)$ levels have $n$ nodes, source and destination locate the first level and the last level, respectively.

\begin{figure}[t]%
\centering
\includegraphics[scale=0.38]{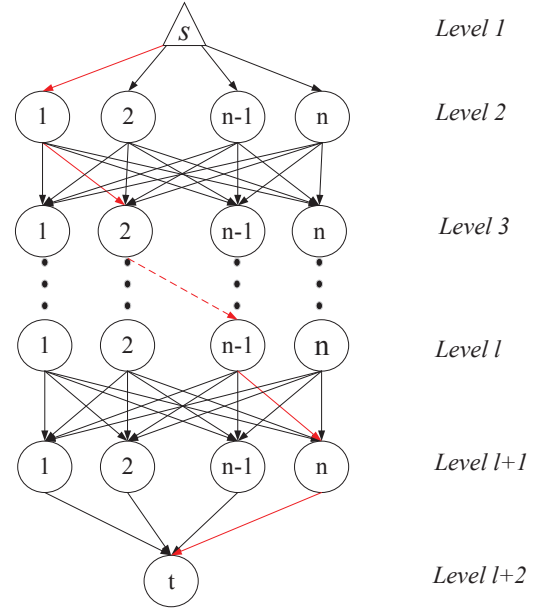}
\caption{The multilevel network}%
\label{multilevelnetwork}%
\end{figure}

Fig. \ref{multilevelnetwork} depicts a multilevel network, the nodes at the top level has arcs to all nodes in the next level. Because of the topology characteristics, the path from source $s$ to destination $t$ traverses $l+2$ nodes properly, including two end nodes. Each arc has a two-tuple attribute, including the cost and delay. Since the upper delay bound is $\Delta$, we set that each arc between neighbor levels has the delay upper bound $\frac{\Delta-l*max\{d_1,...,d_n\}}{(l+1)}$. In fact, the arcs in the multilevel network denote the corresponding minimum cost path between the two end nodes in the original network. That is, when constructing the multilevel network, we need first to find the minimum cost path, between any pair of nodes in the original network, which satisfies the end-to-end delay constraint. These paths are called delay-guaranteed shortest paths in the definition.
\begin{definition}
Given two nodes $\{s,t\}$ and a delay upper bound $\Delta$, the \emph{delay-guaranteed shortest path} is the minimum cost path to connect $s$ and $t$ at the same time the end-to-end delay is less or equal to $\Delta$.
\end{definition}

From the proof of theorem \ref{NPproof}, we can easily conclude that finding the delay-guaranteed shortest path is also a NP-hard problem. However, if we assume $\Delta$ is a bounded integer, then we can use a dynamic programming approach similar to Floyd's shortest path algorithm \cite{Floyd1969Algorithm,Kompella1993Multicast}. We use $\gamma_d(u,v)$ to denote the cost of the shortest path from $u$ to $v$ with the end-to-end delay being exactly $d$. If multiple delay-guaranteed shortest paths exhibit the same cost, then the one with the least delay is chosen. $\psi_C(u,v)$ and $\psi_D(u,v)$ denote the cost and the delay of the delay-guaranteed shortest path, respectively. The formulations of $\gamma_d(u,v)$ and $\psi_C(u,v)$ are as follows:

\begin{align}
 \gamma_d(u,v)&= \underset{w\in V}{\text{min}}\quad\{\gamma_{d-D(w,v)}(u,w)+C(w,v)\} \label{dynamic1} \\
 \psi_C(u,v) & = \underset{d< \delta}{\text{min}}\quad\gamma_d(u,v) \label{dynamic2}
\end{align}

The weight of each arc is the sum of two parts, and the corresponding delay-guaranteed shortest path is the first part. The other part is the weight of node. Take the red link between the level 2 and level 3 for example, we first find the delay-guaranteed shortest path with the delay upper bound $\frac{\Delta-l*max\{d_1,...,d_n\}}{(l+1)}$, and we can get $\psi_C(1,2)$,$\psi_D(1,2)$. The cost and delay of node $2$ are $c_2$ and $d_2$. The final two-tuple weight of the arc is $(\psi_C(1,2)+c_2,\psi_D(1,2)+d_2)$.

\begin{theorem}
\label{feasibleproof}
In the multilevel network, the end-to-end delay of any path connecting $s$ and $t$ is not larger than the upper bound $\Delta$.
\end{theorem}

\begin{IEEEproof}
The two-tuple weight of the arcs are the sum of two parts. The first part is the delay of the concerned delay-guaranteed shortest paths, and this delay is less than $\frac{\Delta-l*max\{d_1,...,d_n\}}{(l+1)}$, then the total sum of the first part is $\frac{\Delta-l*max\{d_1,...,d_n\}}{(l+1)}*l<\Delta-l*max\{d_1,...,d_n\}$. The second part is the delay of the concerned nodes, and the sum is $l*max\{d_1,...,d_n\}$. We can easily conclude that the sum delay of the two parts is then less than $\Delta$. Thus, Theorem \ref{feasibleproof} is proved.
\end{IEEEproof}

Though Theorem \ref{feasibleproof} ensures that any path can be feasible as far as the end-to-end delay constraint. However, some nodes may occur several times in different levels, which means that they are required to deploy multiple VNFs. In Section \ref{introduction}, we have assumed that each node is permitted to deploy only one VNF. For satisfying this assumption, we use a greedy algorithm to find the expected path. For any selected node $u$ in $i-level$, we set the cost of all arcs incident to $u$ in the bottom levels as $\infty$. The greedy algorithm runs from source, in each iteration, it selects the minimum cost arc to reach the next level and update the cost of all arcs. The detail of our greedy algorithm can be seen in Algorithm \ref{alg:Greedy}.

\begin{algorithm}[h]
\caption{The Greedy Algorithm (GA) }
\label{alg:Greedy}
\small
\begin{algorithmic}[1]
\REQUIRE An undirected network $G\mathrm{=}(V,E,\emph{c(v)},\emph{d(v)},\emph{c(e)},\emph{d(e)})$ and a request  $r\mathrm{=}(s,d,SFC,\Delta)$.
\ENSURE A path $P$ connect $s$ and $d$.
\STATE  $MNetwork=\textbf{Multilevel-Network($G,r$)}$.
\STATE $Currentnode = s$
\STATE  update the two-tuple weights of all arcs incident to $t$.
\FOR {$k\mathrm{=}1$ to $l+1$}
\STATE  update the two-tuple weights of all arcs incident to $Currentnode$ in different levels that lies below $k-level$.
\STATE  select the minimum cost edge $p_k$ connecting $Currentnode$ to the $(k+1)-level$.
\STATE  $P=P+p_k$.
\STATE  update the $Currentnode$ with the other end node of the selected edge in $(k+1)-level$.
\ENDFOR
\STATE   return $P$
\end{algorithmic}
\textbf{Multilevel-Network($G,r$)}
\begin{algorithmic}[1]
\STATE  $l=|SFC|$, the multilevel network has $l+2$ levels, and the delay of neighbor levels is $\Delta_1=\frac{\Delta-l*max\{d_1,...,d_n\}}{(l+1)}$.
\STATE  Use the dynamic programming in equation \ref{dynamic1} and \ref{dynamic2} with $r$ and $\Delta_1$ to find all the delay-guaranteed shortest paths $\psi(u,v)$ between all pair of nodes.
\STATE  Construct the multilevel network $MNetwork$, using $\psi_C(u,v)$ and $\psi_D(u,v)$ to attach weights.
\STATE Return $MNetwork$.
\end{algorithmic}
\end{algorithm}

\begin{figure*}[t]
\begin{minipage}[t]{0.33\textwidth}
\centering
\setcaptionwidth{0.8\textwidth}
\includegraphics[scale=0.3]{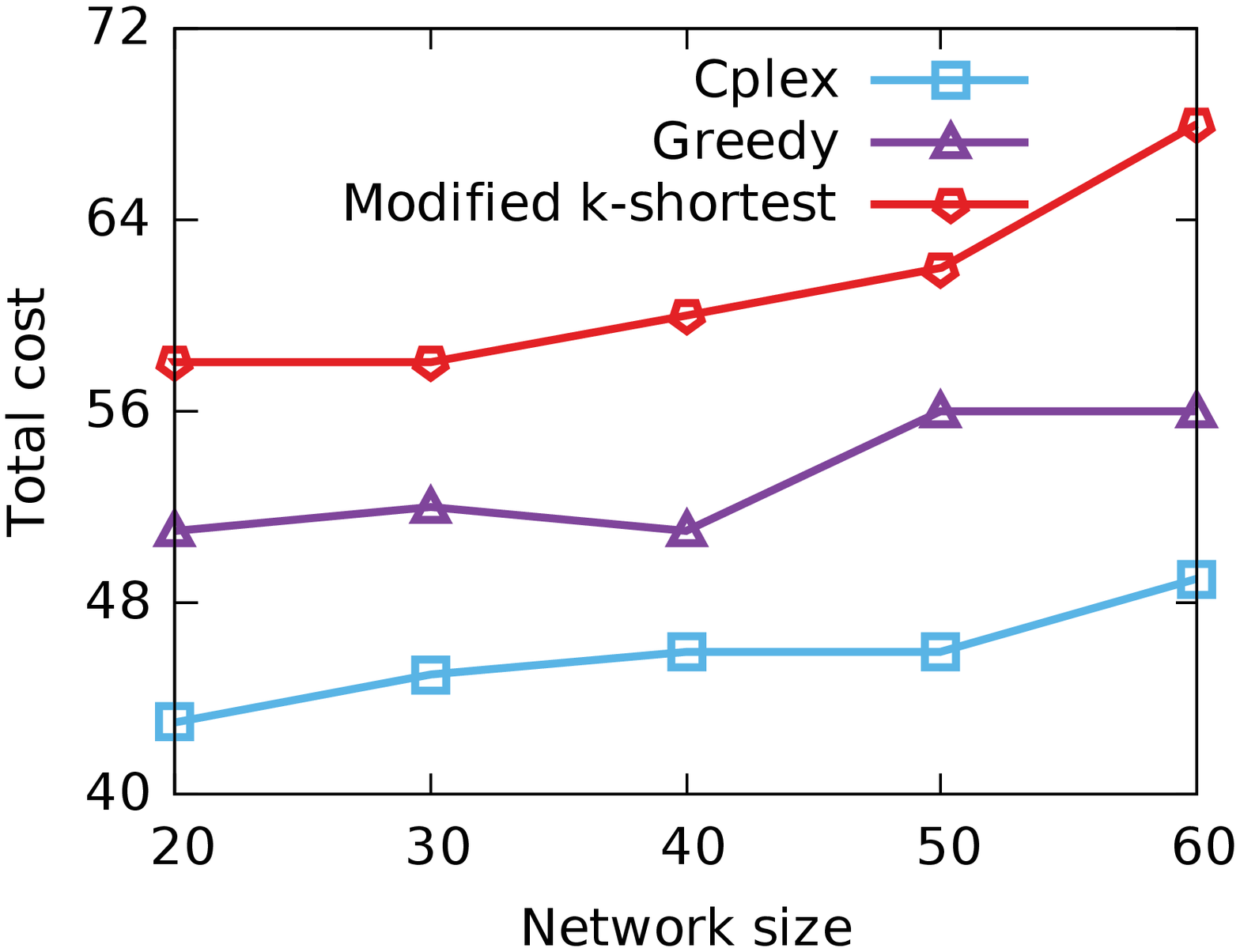}
\caption{The average cost of embedding the SFC with different network sizes.}
\label{averagecost-network}
\end{minipage}%
\begin{minipage}[t]{0.33\textwidth}
\centering
\setcaptionwidth{0.8\textwidth}
\includegraphics[scale=0.3]{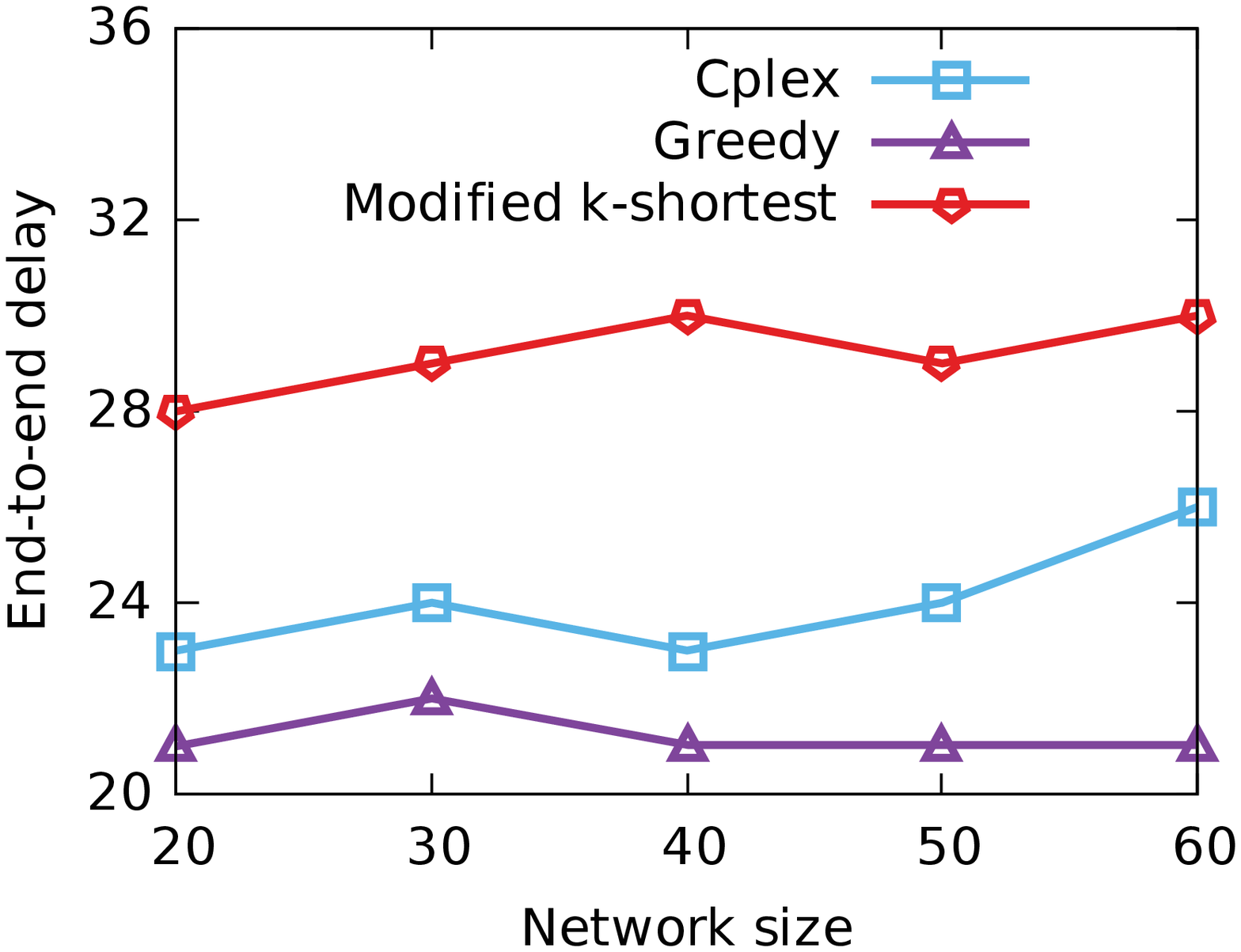}
\caption{The average end-to-end delay of embedding the SFC with different network sizes.}
\label{averagedelay-network}
\end{minipage}%
\begin{minipage}[t]{0.33\textwidth}
\centering
\setcaptionwidth{0.8\textwidth}
\includegraphics[scale=0.3]{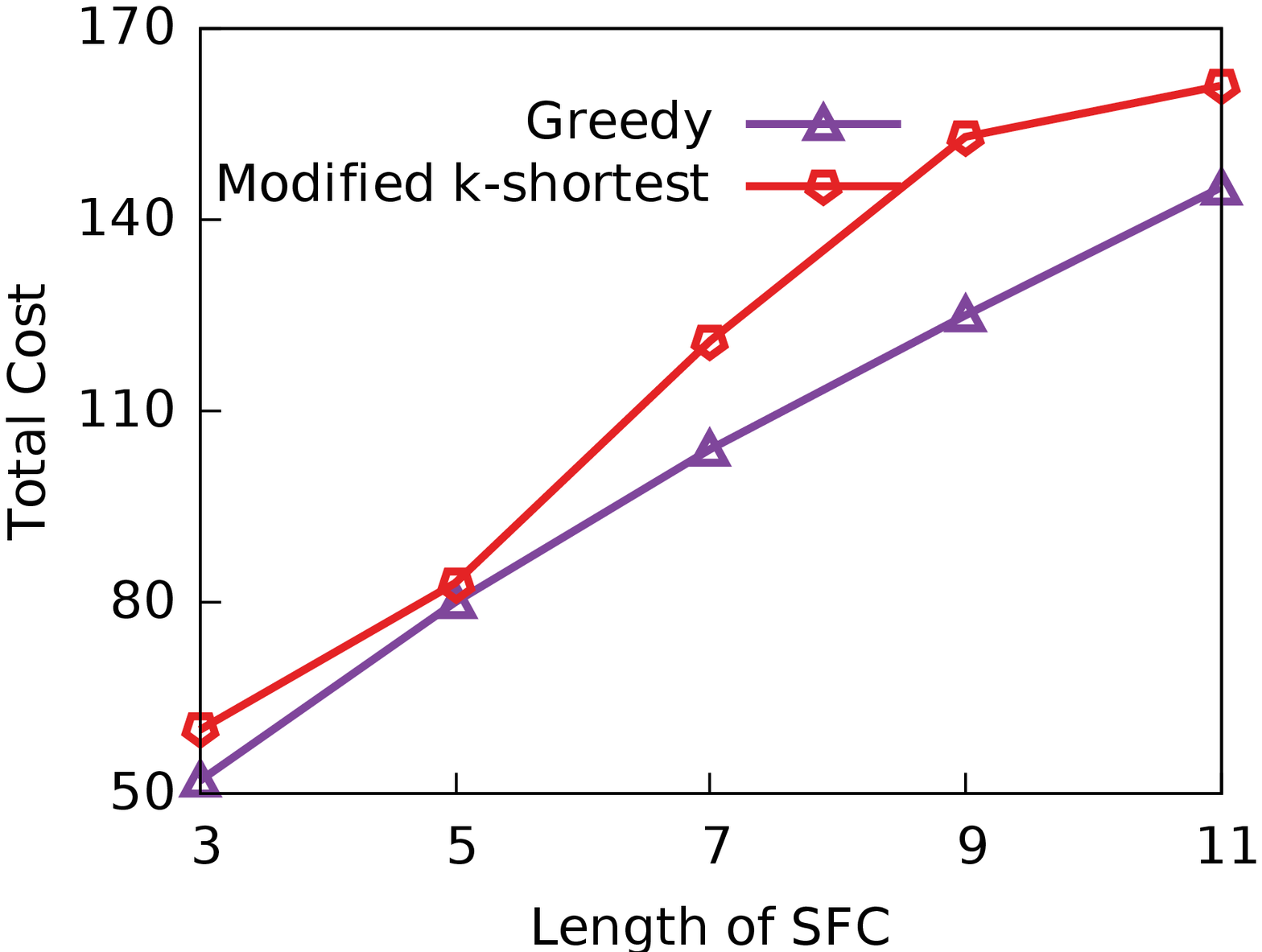}
\caption{The average cost of embedding the SFC with different lengths.}
\label{SFClengthcost}
\end{minipage}%
\end{figure*}

\begin{theorem}
\label{timecomplexity}
The time complexity of Algorithm \ref{alg:Greedy} is $O(\frac{n^3\times\Delta}{l-1})$.
\end{theorem}

\begin{IEEEproof}
When computing the delay-guaranteed shortest paths, we take similar computation loops over all pair of nodes and over all intermediate nodes like Floyd's shortest path algorithm, whose time complexity is $O(n^3)$. Additionally, Equations \ref{dynamic1} and \ref{dynamic2} calculate the loops over all possible values of delay from 1 to $(\frac{\Delta}{l-1}-1)$, then the time complexity of construction the multilevel network is $O(\frac{n^3\times\Delta}{l-1})+O(l\times n^2)=O(\frac{n^3\times\Delta}{l-1})$. After constructing the multilevel network, the path selection in each iteration needs to compare $n$ times. Thus, the time complexity of finding paths in the multilevel network is $O(l\times n)$. The total time complexity of the algorithm is $O(\frac{n^3\times\Delta}{l-1})+O(l\times n)=O(\frac{n^3\times\Delta}{l-1})$. Thus, Theorem \ref{timecomplexity} is proved.
\end{IEEEproof}

\section{NUMERICAL RESULTS}

In this section, we present numerical results of the greedy algorithm. As a comparison, we compare our greedy algorithm to a modified k-shortest paths algorithm \cite{Yen1971FINDING}. The modified k-shortest algorithm first finds k shortest paths between $s$ and $t$, and then select the minimum cost path which satisfies the constraints of our D-MCS model. Also, we use the CPLEX \cite{ibmCplex} to find the optimum solution with small-scale experiments.

\begin{table}[t]
\caption{The topology characteristics of the network. MSCP and MSDP mean the average value of the minimum cost paths and the minimum delay paths for all pairs of nodes, respectively.}
\label{topologycharacterstic} \centering
\begin{tabular}{|c|c|c|}
  \hline
  % after \\: \hline or \cline{col1-col2} \cline{col3-col4} ...
    \# of nodes& MSCP & MSDP \\
   \hline
 20 & 26.3 & 9.2\\
 30 & 27 & 9.9\\
 40 & 27.23 & 10.1\\
 50 & 28.28 & 10.3\\
 60 & 28.32 & 10.5\\
  \hline
\end{tabular}
\end{table}

We first generate synthetic networks using the ER random graphs, with node number (servers) ranging from 20 to 60. All nodes of the network are placed in a rectangle with grid size $[20\times 20]$, the link connection cost between two nodes is set to their Euclidean distance and the delay of each edge is selected randomly from $[1,5]$. The cost and delay of nodes are selected from $[4,8]$ and $[1,3]$, respectively. Before evaluating the algorithms, we first observe the topology characteristics of our synthetic random network. As shown in Table \ref{topologycharacterstic}, as the number of nodes grows up, the average minimum cost path and the average minimum delay path between each pair nodes are almost fixed.

In our experiments, we select the source and destination randomly. We first study the impact of network size. Fig. \ref{averagecost-network} and Fig. \ref{averagedelay-network} depicts the variations of the total cost and the end-to-end delay as the growth of network size. Firstly, the total cost and the end-to-end delay of the embedding SFC is not proportioned to the network size. This can be explained by the Table \ref{topologycharacterstic}, which networks with different sizes in our experiments have almost the same topology characteristics as far as the average minimum delay path and the average minimum cost path. In Fig. \ref{averagecost-network}, the curve of greedy algorithm lies between that of the modified k-shortest paths algorithm and the optimization solution. The total cost of the greedy algorithm can be reduced 12.94\% in average compared to the modified k-shortest algorithm. Although the optimization solutions have lower total cost, we can see from Fig. \ref{averagedelay-network} that the solutions of the greedy algorithm have the lowest end-to-end delay. This phenomenon satisfies our observation, which a stricter end-to-end delay constraint leads to a higher total cost.

We additionally study the impact of the SFC length. In this experiment, we set the network size as 500. Accordingly, this scale is not proper to use CPLEX to solve. In Fig. \ref{SFClengthcost}, we can easily conclude that the total cost of the embedding SFC increases as the growth of length. This is because that the longer of SFC will occupy more links. Indeed, the average number of hops to connect any pair of nodes is $\frac{ln N}{ln <k>}=\frac{ln N}{pN}$, in which $N$ denotes the number of nodes and $p$ denotes the probability of connecting two nodes. This characteristic decides that the SFC length plays a key role in the length of the path connecting source and destination.

\begin{comment}
\begin{figure}[t]%
\centering
\includegraphics[scale=0.38]{graph/averagecost_network}
\caption{The average cost of the SFC with different network size}%
\label{averagecost-network}%
\end{figure}

\begin{figure}[t]%
\centering
\includegraphics[scale=0.38]{graph/averagedelay_network}
\caption{The average end-to-end delay of the SFC with different network size}%
\label{averagedelay-network}%
\end{figure}

\begin{figure}[t]%
\centering
\includegraphics[scale=0.38]{graph/SFClengthcost}
\caption{The average cost of the SFC with different length of the SFC}%
\label{SFClengthcost}%
\end{figure}
\end{comment}

\section{CONCLUSION}
In this paper, we have studied the delay-guaranteed minimum cost embedding SFC for unicast. Our main contributions are as follows: (i) the embedding path for the unicast must traverse at least $l$ nodes (Fig. \ref{exampleshow}), where $l$ denotes the length of the SFC; (ii) we have proved the NP-hardness of the problem through reducing it to the classical KNAPSACK problem (Fig. \ref{graphinstance}); (iii) we have designed a greedy algorithm in the multilevel network (Fig. \ref{multilevelnetwork}) to solve the NP-hard problem, and do simulations to check the validity of the designed algorithm (Fig. \ref{averagecost-network}-\ref{SFClengthcost}). Future work may include extensions that applied to multicast with end-to-end delay variations and comparisons with experimental results.

\bibliographystyle{IEEEtran}
% argument is your BibTeX string definitions and bibliography database(s)
\bibliography{SFCembedding}

% Generated by IEEEtran.bst, version: 1.14 (2015/08/26)
\begin{thebibliography}{10}
\providecommand{\url}[1]{#1}
\csname url@samestyle\endcsname
\providecommand{\newblock}{\relax}
\providecommand{\bibinfo}[2]{#2}
\providecommand{\BIBentrySTDinterwordspacing}{\spaceskip=0pt\relax}
\providecommand{\BIBentryALTinterwordstretchfactor}{4}
\providecommand{\BIBentryALTinterwordspacing}{\spaceskip=\fontdimen2\font plus
\BIBentryALTinterwordstretchfactor\fontdimen3\font minus
  \fontdimen4\font\relax}
\providecommand{\BIBforeignlanguage}[2]{{%
\expandafter\ifx\csname l@#1\endcsname\relax
\typeout{** WARNING: IEEEtran.bst: No hyphenation pattern has been}%
\typeout{** loaded for the language `#1'. Using the pattern for}%
\typeout{** the default language instead.}%
\else
\language=\csname l@#1\endcsname
\fi
#2}}
\providecommand{\BIBdecl}{\relax}
\BIBdecl

\bibitem{gerdfeldter2015setting}
A.~GERDFELDTER, P.~HIGGS, P.~LJUNGBERG, N.~MITRA, and M.~PERSSON, ``Setting the
  future media services architecture,'' \emph{ERICSSON REVIEW}, 2015.

\bibitem{ren2018minimum}
B.~Ren, G.~Cheng, and D.~Guo, ``Minimum-cost forest for uncertain multicast
  with delay constraints,'' \emph{Tsinghua Science and Technology}, vol.~24,
  no.~2, pp. 157--159, 2018.

\bibitem{anvato}
``Anvato,'' \url{http://www.anvato.com/}.

\bibitem{NFVsurvey}
B.~Han, V.~Gopalakrishnan, L.~Ji, and S.~Lee, ``Network function
  virtualization: Challenges and opportunities for innovations,'' \emph{{IEEE}
  Communications Magazine}, vol.~53, no.~2, pp. 90--97, 2015.

\bibitem{Netflix}
``Netflix,'' \url{http://www.netflix.com/}.

\bibitem{HanGJL15}
B.~Han, V.~Gopalakrishnan, L.~Ji, and S.~Lee, ``Network function
  virtualization: Challenges and opportunities for innovations,'' \emph{{IEEE}
  Communications Magazine}, vol.~53, no.~2, pp. 90--97, 2015.

\bibitem{Karp1975Reducibility}
Karp and R.~M, ``Reducibility among combinatorial problems,'' \emph{Journal of
  Symbolic Logic}, vol.~40, no.~4, pp. 618--619, 1975.

\bibitem{Floyd1969Algorithm}
R.~W. Floyd, ``Algorithm 97: Shortest path,'' \emph{Communications of the Acm},
  vol.~5, no.~6, p. 345, 1969.

\bibitem{Kompella1993Multicast}
V.~P. Kompella, J.~C. Pasquale, and G.~C. Polyzos, ``Multicast routing for
  multimedia communication,'' \emph{IEEE/ACM Transactions on Networking},
  vol.~1, no.~3, pp. 286--292, 1993.

\bibitem{Yen1971FINDING}
J.~Y. Yen, ``Finding the k shortest loopless paths in a network,''
  \emph{Management Science}, vol.~17, no.~11, pp. 712--716, 1971.

\bibitem{ibmCplex}
``Cplex,'' \url{https://www.ibm.com/products/ilog-cplex-optimization-studio}.

\end{thebibliography}

%
%\begin{thebibliography}{1}
%
%\bibitem{IEEEhowto:kopka}
%H.~Kopka and P.~W. Daly, \emph{A Guide to \LaTeX}, 3rd~ed.\hskip 1em plus
%  0.5em minus 0.4em\relax Harlow, England: Addison-Wesley, 1999.
%
%\end{thebibliography}
%
%% biography section
%%
%% If you have an EPS/PDF photo (graphicx package needed) extra braces are
%% needed around the contents of the optional argument to biography to prevent
%% the LaTeX parser from getting confused when it sees the complicated
%% \includegraphics command within an optional argument. (You could create
%% your own custom macro containing the \includegraphics command to make things
%% simpler here.)
%%\begin{IEEEbiography}[{\includegraphics[width=1in,height=1.25in,clip,keepaspectratio]{mshell}}]{Michael Shell}
%% or if you just want to reserve a space for a photo:
%
%\begin{IEEEbiography}{Michael Shell}
%Biography text here.
%\end{IEEEbiography}
%
%% if you will not have a photo at all:
%\begin{IEEEbiographynophoto}{John Doe}
%Biography text here.
%\end{IEEEbiographynophoto}
%
%% insert where needed to balance the two columns on the last page with
%% biographies
%%\newpage
%
%\begin{IEEEbiographynophoto}{Jane Doe}
%Biography text here.
%\end{IEEEbiographynophoto}
%
%% You can push biographies down or up by placing
%% a \vfill before or after them. The appropriate
%% use of \vfill depends on what kind of text is
%% on the last page and whether or not the columns
%% are being equalized.
%
%%\vfill
%
%% Can be used to pull up biographies so that the bottom of the last one
%% is flush with the other column.
%%\enlargethispage{-5in}
%
\end{document}